\documentclass[twocolumn,showpacs,amsfonts,amsmath,amssymb,aps,pra]{revtex4}
\usepackage{bm}
\usepackage{amsfonts}

\begin{document}
\hbadness=10000

 To be published in Phys. Rev. A
 \vspace{1cm}

\title{Center-of-mass motion in many-body theory of BEC}
\author{Iwo Bialynicki-Birula}
 \email{birula@cft.edu.pl}
\affiliation{Center for Theoretical Physics, Polish Academy of Sciences,\\
Lotnikow 32/46, 02-668 Warsaw, Poland\\ and Institute of Theoretical Physics, Warsaw
University}
\author{Zofia Bialynicka-Birula}
\affiliation{Institute of Physics, Polish Academy of Sciences and College of Science,\\
Al. Lotnik\'ow 32/46, 02-668 Warsaw, Poland}

\begin{abstract}

The method of generating a family of new solutions starting from any wave
function satisfying the nonlinear Schr\"odinger equation in a harmonic
potential proposed recently [ J. J. Garc\'{\i}a-Ripoll, V. M.
P\'erez-Garc\'{\i}a, and V. Vekslerchik, Phys. Rev. E {\bf 64}, 056602 (2001)]
is extended to many-body theory of mutually interacting particles. Our method
is based on a generalization of the displacement operator known in quantum
optics and results in the separation of the center of mass motion from the
internal dynamics of many-body systems. The center of mass motion is analyzed
for an anisotropic rotating trap and a region of instability for intermediate
rotational velocities is predicted.
\end{abstract}
\pacs{05.30.Jp,03.75.Fi,03.65.-w}

\maketitle

\section{Introduction}

In a recent paper Garc\'{\i}a-Ripoll, P\'erez-Garc\'{\i}a, and Vekslerchik
\cite{gpv} have studied the properties of the solution of the nonlinear
Schr\"odinger equation in a time-dependent, anisotropic harmonic potential
\begin{equation}\label{nonleq}
 i\partial_t\psi({\mathbf r}, t)
 = \left(-\frac{1}{2}\Delta +
 \frac{1}{2}{\mathbf r}\!\cdot\!{\hat A}(t)\!\cdot\!{\mathbf r} +
 G(\vert\psi({\mathbf r}, t)\vert)\right)\psi({\mathbf r}, t),
\end{equation}
where ${\hat A}(t)$ is an arbitrary symmetric positive $3\times 3$ matrix and $G$ is an
arbitrary real function. They have shown that from every solution of this equation one
may obtain the whole family of new solutions by a translation accompanied by a change of
the phase
\begin{eqnarray}\label{oldnew}
 \psi({\mathbf r}, t)
 \to \psi'({\mathbf r}, t) = \psi({\mathbf r} - {\mathbf R}(t), t)e^{i\theta({\mathbf r}, t)},
\end{eqnarray}
provided ${\mathbf R}(t)$ is a solution of the classical equation of motion in
the harmonic potential
\begin{eqnarray}\label{eqmotion}
 \frac{d^2{\mathbf R}(t)}{dt^2} = -{\hat A}(t)\!\cdot\!{\mathbf R}(t)
\end{eqnarray}
and the phase $\theta({\mathbf r}, t)$ is given by the formula
\begin{eqnarray}\label{theta}
 \theta({\mathbf r}, t) = {\mathbf r}\!\cdot\!\frac{d{\mathbf R}(t)}{dt} - f(t),
\end{eqnarray}
where $f(t)$ is the classical action calculated along the trajectory ${\mathbf
R}(t)$ (a factor of $-1/2$ is missing in Eq.~(12) of Ref.~\cite{gpv})
\begin{eqnarray}\label{action}
f(t) = \frac{1}{2}\int_0^t\!\!dt\left(\frac{d{\mathbf
R}(t)}{dt}\!\cdot\!\frac{d{\mathbf R}(t)}{dt} - {\mathbf R}(t)\!\cdot\!{\hat
A}(t)\!\cdot\!{\mathbf R}(t)\right).
\end{eqnarray}
This result implies that the motion of the center of the wave packet is
governed by the classical equation (\ref{eqmotion}) and it decouples from the
internal motion that determines the shape of the wave packet. It has been
argued in \cite{gpv} that such a decoupling may be of significance for the
dynamics of Bose-Einstein condensates, since they are often described in the
mean-field approximation by the nonlinear Schr\"odinger equation.

In this paper we extend the conclusions of Ref.~\cite{gpv} to the general case of a {\em
mutually interacting} many-body quantum system. Our generalization is inspired by the
observation that the new solutions (\ref{oldnew}) may be generated from the initial
solution by the displacement operator well known in quantum optics \cite{glauber,scully}
\begin{eqnarray}\label{displacement}
 e^{i({\hat{\mathbf r}}\cdot{\mathbf P}(t)-{\hat{\mathbf p}}\cdot{\mathbf R}(t))},
\end{eqnarray}
where ${\hat{\mathbf r}}$ and ${\hat{\mathbf p}} = -i\nabla$ are the quantum
mechanical position and momentum operators whereas ${\mathbf R}(t)$ and
${\mathbf P}(t)$ represent the phase-space trajectory of a classical particle
in the harmonic potential. In order to show that the transformation
\begin{eqnarray}\label{newsoln}
 \psi({\mathbf r}, t) \to \psi'({\mathbf r}, t) =
 e^{i({\hat{\mathbf r}}\cdot{\mathbf P}(t)-{\hat{\mathbf p}}\cdot{\mathbf R}(t))}
 \psi({\mathbf r}, t)
\end{eqnarray}
is the same as that given by (\ref{oldnew}), we apply the Baker-Hausdorff
formula (using the canonical commutation relations between ${\hat{\mathbf r}}$
and ${\hat{\mathbf p}}$) to write the displacement operator in the form
\begin{eqnarray}\label{bh}
 e^{i({\hat{\mathbf r}}\cdot{\mathbf P}(t)-{\hat{\mathbf p}}\cdot{\mathbf R}(t))}
 = e^{-i{\mathbf R}(t)\cdot{\mathbf P}(t)/2}
 e^{i{\hat{\mathbf r}}\cdot{\mathbf P}(t)}
 e^{-i{\hat{\mathbf p}}\cdot{\mathbf R}(t)}.
\end{eqnarray}
The last term on the r.h.s shifts the argument of the wave function by
$-{\mathbf R}(t)$ and the middle term produces the first term of the phase
(\ref{theta}). Finally, we note that the action $f(t)$ after integration by
parts reduces to the boundary terms
\begin{eqnarray}\label{boundary}
 f(t) = \left.\frac{1}{2}\frac{d{\mathbf R}(t)}{dt}\!\cdot\!{\mathbf
 R}(t)\right\vert^t_0.
\end{eqnarray}
Therefore, the first term on the r.h.s. of (\ref{bh}) reproduces the second
term of the phase (\ref{theta}) (up to an irrelevant constant phase given by
the lower boundary term).

The action of the displacement operator on a solution of a nonlinear
Schr\"odinger equation leads to a shift of the center of the wave packet. The
vector ${\mathbf R}(t)$ that determines this shift is a solution of the
classical equations of motion in the harmonic potential. Therefore, the center
of the wave packet always follows a classical trajectory. In the present paper,
we cary out a similar analysis for an interacting many-body system. The role of
the center of the wave packet is played by the center of mass operator of the
whole system.

In Section II we construct the transformation analogous to (\ref{newsoln}) in
the many-body theory of a system of mutually interacting bosons (or fermions)
and we show that it also transforms solutions of the Schr\"odinger equation
into new solutions of this equation. In Section III we show that all these
properties are a result of a complete decoupling of the center of mass dynamics
from the internal dynamics of the system confined in a harmonic potential. This
decoupling has been noticed before (cf., for example, Ref.~\cite{pp}). In
Section IV we analyze the spectrum of stationary states for a rotating trap and
we point out that the decoupling causes the splitting of the spectrum. We also
show that for an anisotropic trap there always exists a region of instability
at intermediate rotational velocities.

\section{Displacement operator in many-body theory}

The many-body theory of interacting atoms will be described within the
formalism of second quantization. We shall use systematically the Schr\"odinger
picture --- all operators will be time independent. The Hamiltonian of the
system of atoms contained in a harmonic trap (in natural units, $m = 1,\;\hbar
= 1$) has the form
\begin{eqnarray}\label{ham}
&&{\hat H}(t) = \frac{1}{2}\int\!\!d^3r{\hat\psi}^\dagger({\mathbf
r})\left(-{\bm\nabla}\!\cdot\!{\bm\nabla} + {\mathbf r}\!\cdot\!{\hat
A}(t)\!\cdot\!{\mathbf r}\right) {\hat\psi}({\mathbf r})\nonumber\\
 &+&\frac{1}{2}\int\!\!d^3r\!\int\!\!d^3r'{\hat\psi}^\dagger({\mathbf
r}){\hat\psi}^\dagger({\mathbf r}') V({\mathbf r}-{\mathbf
r}'){\hat\psi}({\mathbf r}'){\hat\psi}({\mathbf r}),
\end{eqnarray}
where ${\hat A}(t)$ is the matrix of the harmonic potential of the previous
Section and $V({\mathbf r}-{\mathbf r}')$ is an arbitrary two-particle
interaction potential. The field operators ${\hat\psi}({\mathbf r})$ and
${\hat\psi}^\dagger({\mathbf r})$ that annihilate and create particles at the
point ${\mathbf r}$ obey the standard Bose-Einstein or Fermi-Dirac commutation
relations (we disregard the spin degrees of freedom)
\begin{eqnarray}\label{cr}
\left[{\hat\psi}^\dagger({\mathbf r}),{\hat\psi}({\mathbf r}')\right]_{\mp} =
\delta^{(3)}({\mathbf r}-{\mathbf r}').
\end{eqnarray}
In the construction of the displacement operator in many-body theory (following
our earlier works \cite{bb,bb1}) we employ the operators of the total number of
particles ${\hat N}$, of the of the total position ${\hat{\mathbf R}}$, and of
the total momentum ${\hat{\mathbf P}}$:
\begin{subequations}\label{ops}
\begin{eqnarray}
 {\hat N} &=& \int\!\!d^3r{\hat\psi}^\dagger({\mathbf r}){\hat\psi}
 ({\mathbf r}),\label{n}\\
 {\hat{\mathbf R}}
 &=& \int\!\!d^3r{\hat\psi}^\dagger({\mathbf r}){\mathbf r}{\hat\psi}({\mathbf r}),\label{r}\\
 {\hat{\mathbf P}}
 &=& -i\int\!\!d^3r{\hat\psi}^\dagger({\mathbf r}){\bm\nabla}{\hat\psi}({\mathbf r}).\label{p}
\end{eqnarray}
\end{subequations}
The number of particles operator ${\hat N}$ is a constant of motion and it
commutes with ${\hat{\mathbf R}}$ and ${\hat{\mathbf P}}$. The operators
${\hat{\mathbf R}}$ and ${\hat{\mathbf P}}$ satisfy the commutation relations
\begin{eqnarray}\label{cr1}
 \left[{\hat{R}_k}, {\hat{P}_l}\right] = i{\hat N}\delta_{kl}.
\end{eqnarray}

We define the many-body unitary displacement operator ${\hat D}(t)$ by the same
general formula (\ref{displacement}) but with second-quantized operators
${\hat{\mathbf R}}$ and ${\hat{\mathbf P}}$  playing now the role of position
and momentum,
\begin{eqnarray}\label{displacement1}
 {\hat D}(t)
 = e^{i({\hat{\mathbf R}}\cdot{\mathbf P}(t)-{\hat{\mathbf P}}\cdot{\mathbf R}(t))}.
\end{eqnarray}
We would like to note that the new displacement operator ${\hat D}(t)$ does not
share {\em all} the properties of the Glauber displacement operator in quantum
optics. Namely, it does not generate a coherent state when acting on the vacuum
state but leaves the vacuum state unchanged ${\hat
D}(t)\vert0\rangle=\vert0\rangle$. This is due to the fact that the many-body
vacuum state is not an exact counterpart of the QED vacuum.

In the proof of the invariance of the Schr\"odinger equation
\begin{eqnarray}\label{schrod}
 i\partial_t\vert\Psi(t)\rangle = {\hat H}(t)\vert\Psi(t)\rangle
\end{eqnarray}
under the action of ${\hat D}(t)$ we shall again employ the Baker-Hausdorff
decomposition of the displacement operator:
\begin{eqnarray}\label{bh1}
 e^{i({\hat{\mathbf R}}\cdot{\mathbf P}(t)-{\hat{\mathbf P}}\cdot{\mathbf R}(t))}
 = e^{-i{\hat N}C(t)} e^{i{\hat{\mathbf R}}\cdot{\mathbf P}(t)}
 e^{-i{\hat{\mathbf P}}\cdot{\mathbf R}(t)},
\end{eqnarray}
where $C(t) = {\mathbf R}(t)\!\cdot\!{\mathbf P}(t)/2$. Next we observe that
the three factors on the r.h.s. of this formula act in the following way on the
annihilation operators
\begin{subequations}\label{act}
\begin{eqnarray}
 e^{i{\hat N}C(t)}{\hat\psi}({\mathbf r}) e^{-i{\hat N}C(t)}
 = e^{-iC(t)}{\hat\psi}({\mathbf r}),\\
 e^{-i{\hat{\mathbf R}}\cdot{\mathbf P}(t)}{\hat\psi}({\mathbf r})
 e^{i{\hat{\mathbf R}}\cdot{\mathbf P}(t)}
 = e^{i{\mathbf r}\cdot{\mathbf P}(t)}{\hat\psi}({\mathbf r}),\\
 e^{i{\hat{\mathbf P}}\cdot{\mathbf R}(t)}{\hat\psi}({\mathbf r})
 e^{-i{\hat{\mathbf P}}\cdot{\mathbf R}(t)}
 = {\hat\psi}({\mathbf r} - {\mathbf R}(t)),
\end{eqnarray}
\end{subequations}
where we have used the following commutation relations between the operators
${\hat N}$, ${\hat{\mathbf R}}$, and ${\hat{\mathbf P}}$ and the annihilation
operators,
\begin{subequations}
\begin{eqnarray}
 \left[{\hat{N}}, {\hat\psi}({\mathbf r})\right]
 &=& -{\hat\psi}({\mathbf r}),\\
 \left[{\hat{\mathbf R}}, {\hat\psi}({\mathbf r})\right]
 &=& -{\mathbf r}{\hat\psi}({\mathbf r}),\\
 \left[{\hat{\mathbf P}}, {\hat\psi}({\mathbf r})\right]
 &=& i{\mathbf\nabla}{\hat\psi}({\mathbf r}).
\end{eqnarray}
\end{subequations}
The formulas for the creation operators are obtained by hermitian conjugation.
With the help of Eqs.~(\ref{act}) we may calculate the action of the
displacement operator on the Hamiltonian
\begin{widetext}
\begin{eqnarray}\label{trham}
 &&{\hat D}^\dagger(t){\hat H}(t){\hat D}(t)
 =  \frac{1}{2}\int\!\!d^3r{\hat\psi}^\dagger({\mathbf r})\left(-i{\bm\nabla}
 + {\mathbf P}(t)\right)^2{\hat\psi}({\mathbf r})\nonumber\\
 &+&\frac{1}{2}\int\!\!d^3r{\hat\psi}^\dagger({\mathbf r})
 \left({\mathbf r} + {\mathbf R}(t)\right)\!\cdot\!{\hat A}(t)\!\cdot\!
 \left({\mathbf r} + {\mathbf R}(t)\right){\hat\psi}({\mathbf r})
 + \frac{1}{2}\int\!\!d^3r\!\int\!\!d^3r'{\hat\psi}^\dagger({\mathbf r})
 {\hat\psi}^\dagger({\mathbf r}')V({\mathbf r}-{\mathbf r}')
 {\hat\psi}({\mathbf r}'){\hat\psi}({\mathbf r})\nonumber\\
 &=& {\hat H}(t) + {\hat{\mathbf P}}\!\cdot\!{\mathbf P}(t)
 + {\hat{\mathbf R}}\!\cdot\!{\hat A}(t)\!\cdot\!{\mathbf R}(t)
 + \frac{{\hat N}}{2}\left({\mathbf P}(t)\!\cdot\!{\mathbf P}(t)
 + {\mathbf R}(t)\!\cdot\!{\hat A}(t)\!\cdot\!{\mathbf R}(t)\right).
\end{eqnarray}
\end{widetext}
The interaction term in the Hamiltonian does not change because it is invariant
under translation and under the change of the phase of the field operators. The
last ingredient needed to prove the invariance of the Schr\"odinger equation is
the following transformation formula for the time derivative
\begin{eqnarray}\label{trder}
 &&{\hat D}^\dagger(t)i\partial_t{\hat D}(t)
 = i\partial_t + \frac{{\hat N}}{2}
 \frac{d\left({\mathbf R}(t)\!\cdot\!{\mathbf P}(t)\right)}{dt}\nonumber\\
 &-& {\hat{\mathbf R}}\!\cdot\!\frac{d{\mathbf P}(t)}{dt}
 + {\hat{\mathbf P}}\!\cdot\!\frac{d{\mathbf R}(t)}{dt}
 - {\hat N}{\mathbf R}(t)\!\cdot\!\frac{d{\mathbf P}(t)}{dt}.
\end{eqnarray}
All the terms in this formula that contain functions ${\mathbf R}(t)$ and
${\mathbf P}(t)$ cancel with their counterparts in the transformed Hamiltonian
(with the use of the classical equations of motion) and we finally obtain
\begin{eqnarray}\label{trschrod}
 i\partial_t{\hat D}(t)\vert\Psi(t)\rangle
 = {\hat H}(t){\hat D}(t)\vert\Psi(t)\rangle.
\end{eqnarray}
Therefore, for each classical trajectory we obtain a new solution of the many-body
Schr\"odinger equation, as was the case for the nonlinear Schr\"odinger equation.

In order to connect this result with the effective one-particle theory described in the
Introduction, we shall represent the state vectors by the $n$-particle wave functions
$\Psi({\mathbf r}_1,{\mathbf r}_2,\dots,{\mathbf r}_n,t)$,
\begin{eqnarray}\label{wf}
 \Psi({\mathbf r}_1,{\mathbf r}_2,\dots,{\mathbf r}_n,t)
 = \langle 0\vert{\hat\psi}({\mathbf r}_1){\hat\psi}({\mathbf r}_2)\dots
 {\hat\psi}({\mathbf r}_n)\vert\Psi(t)\rangle.
\end{eqnarray}
Using the unitarity of the displacement operator and the invariance of vacuum
state, we obtain (with the help of the formulas (\ref{act})) the following
relation between the the wave functions of the initial $\vert\Psi(t)\rangle$
and the transformed state ${\hat D}(t)\vert\Psi(t)\rangle$
\begin{widetext}
\begin{eqnarray}\label{trwf0}
 \Psi({\mathbf r}_1,{\mathbf r}_2,\dots,{\mathbf r}_n,t) &\to&
 \Psi'({\mathbf r}_1,{\mathbf r}_2,\dots,{\mathbf r}_n,t)
 = \langle 0\vert{\hat\psi}({\mathbf r}_1){\hat\psi}({\mathbf r}_2)\dots
 {\hat\psi}({\mathbf r}_n){\hat D}(t)\vert\Psi(t)\rangle\nonumber\\
 &=& \langle 0\vert{\hat D}(t){\hat D}^\dagger(t){\hat\psi}({\mathbf r}_1)
 {\hat D}(t){\hat D}^\dagger(t)
 {\hat\psi}({\mathbf r}_2){\hat D}(t){\hat D}^\dagger(t)\dots
 \cdots{\hat D}(t){\hat D}^\dagger(t)
 {\hat\psi}({\mathbf r}_n){\hat D}(t)\vert\Psi(t)\rangle\nonumber\\
 &=& e^{i\theta({\mathbf r}_1, t) + i\theta({\mathbf r}_2, t) + \dots
 + i\theta({\mathbf r}_n, t)}
 \Psi({\mathbf r}_1-{\mathbf R}(t),{\mathbf r}_2-{\mathbf R}(t),
 \dots,{\mathbf r}_n-{\mathbf R}(t),t),
\end{eqnarray}
\end{widetext}
where the phase $\theta({\mathbf r}, t)$ is given by the same formula
(\ref{theta}) as in the one-particle theory. Thus, to each solution of the
many-body Schr\"odinger equation there corresponds the whole family of
solutions labeled by the classical trajectories in a harmonic trap. The new
many-body wave functions are obtained from the original wave function by
exactly the same transformation as in the one-particle case: a time-dependent
shift of its arguments and a multiplication by the same phase factors. The
transformation formula (\ref{trwf0}) acquires a particularly simple form in the
special case, when the initial wave function describes a state in which the
center of mass is not entangled with the internal degrees of freedom,
\begin{eqnarray}\label{noent}
 \Psi({\mathbf r}_1,{\mathbf r}_2,\dots,{\mathbf r}_n,t)
 = \Phi({\bm\rho},t)
 \Phi({\bm\xi}_1,{\bm\xi}_2,\dots,{\bm\xi}_{n-1},t),
\end{eqnarray}
where ${\bm\rho} = ({\mathbf r}_1+{\mathbf r}_2+\dots+{\mathbf r}_n)/n$ and
${\bm\xi}_1,{\bm\xi}_2,\dots,{\bm\xi}_{n-1}$ are some relative coordinates,
invariant under the translations ${\mathbf r}_i \to {\mathbf r}_i + {\mathbf
a}$. Then, the transformation (\ref{trwf0}) affects only the center of mass
wave function
\begin{eqnarray}\label{trwf1}
 \Phi({\bm\rho},t) \to e^{i\theta({\bm\rho}, t)}\Phi({\bm\rho}-{\mathbf
 R}(t),t),
\end{eqnarray}
leaving the wave function of the internal motion intact. A general many-body
wave function can always be written as a sum of product wave functions
(\ref{noent}). The action of the displacement operator on a general wave
function describing a state in which the center of mass motion is entangled
with the internal motion gives
\begin{eqnarray}\label{general}
 &{\hat D}(t)\Psi({\mathbf r}_1,{\mathbf r}_2,\dots,{\mathbf r}_n,t)
 = e^{i\theta({\bm\rho}, t)}&\nonumber\\
 &\times\sum_k\Phi_k({\bm\rho}-{\mathbf R}(t),t)
 \Phi_k({\bm\xi}_1,{\bm\xi}_2,\dots,{\bm\xi}_{n-1},t)&.
\end{eqnarray}
In this case the integration over the center of mass position (either
${\bm\rho}$ or ${\mathbf R}(t)$) introduced in Ref.~\cite{pp} leads to a mixed
internal state of the condensate. The results presented in this Section are
closely related to the fact that, as is shown in the next Section, the total
Hamiltonian (\ref{ham}) may be split into the Hamiltonian of the center of mass
and the Hamiltonian of the internal motion.

\section{Decoupling of the center of mass motion}

The separability of the center of mass dynamics from the internal dynamics in a
harmonic trap follows directly from the form (\ref{ham}) of the Hamiltonian.
Indeed, this Hamiltonian may be written as a sum of two commuting parts: the
Hamiltonian of the center of mass motion ${\hat H}_{\rm CM}(t)$ and the
Hamiltonian ${\hat H}_{\rm I}(t)$ describing the internal structure of the
condensate,
\begin{eqnarray}
 {\hat H}(t) &=& {\hat H}_{\rm CM}(t) + {\hat H}_{\rm I}(t),\label{ham12}\\
 {\hat H}_{\rm CM}(t)
 &=& \frac{{\hat{\mathbf P}}\!\cdot\!{\hat{\mathbf P}}}{2{\hat{N}}}
 + \frac{{\hat{\mathbf R}}\!\cdot\!{\hat{A}(t)}\!\cdot\!
 {\hat{\mathbf R}}}{2{\hat{N}}},\label{hamcm}\\
 {\hat H}_{\rm I}(t) &=& \frac{1}{2}\!\int\!\!d^3r{\hat\psi}^\dagger({\mathbf
 r})\!
 \left(-{\mathbf\nabla}_{\rm C}\!\cdot\!{\mathbf\nabla}_{\rm C}
 + {\mathbf r}_{\rm C}\!\cdot\!
 {\hat A}(t)\!\cdot\!{\mathbf r}_{\rm C}\right)\!{\hat\psi}({\mathbf
 r})\nonumber\\
 + \frac{1}{2}\!\!\!\!\!\!&&\!\!\!\!\!\!\int\!\!d^3r\!\!\int\!\!d^3r'{\hat\psi}^\dagger({\mathbf r})
 {\hat\psi}^\dagger({\mathbf r}') V({\mathbf r}-{\mathbf r}')
 {\hat\psi}({\mathbf r}'){\hat\psi}({\mathbf r}),
\end{eqnarray}
where ${\mathbf\nabla}_{\rm C}$ and ${\mathbf r}_{\rm C}$ are the one-particle
operators shifted by the (normalized) center of mass operators
\begin{eqnarray}\label{shifted}
 {\mathbf r}_{\rm C} = {\mathbf r} - {\hat{\mathbf R}}/{\hat{N}},\;\;
 -i{\mathbf\nabla}_{\rm C} = -i{\mathbf\nabla} - {\hat{\mathbf P}}/{\hat{N}}.
\end{eqnarray}
The sum (\ref{ham12}) describes the dynamics of two independent systems since
the center of mass operators ${\hat{\mathbf R}}$ and ${\hat{\mathbf P}}$
commute with the Hamiltonian ${\hat H}_{\rm I}(t)$, despite the appearance of
these operators in ${\hat H}_{\rm I}(t)$. To prove this statement we observe
that
\begin{eqnarray}\label{proof}
 \left[{\hat{\mathbf R}},{\hat H}_{\rm I}(t)\right]
 \!=\! \left[{\hat{\mathbf R}},{\hat H}(t)\right]
 \!-\! \left[{\hat{\mathbf R}},{\hat H}_{\rm CM}(t)\right]
 \!=\! {\hat{\mathbf P}}\!-\! {\hat{\mathbf P}} \!=\! 0.
\end{eqnarray}
Similarly one can show that $\left[{\hat{\mathbf P}},{\hat H}_{\rm I}(t)\right]
= 0$.

The many-body displacement operator (\ref{displacement1}), constructed from the
center of mass operators ${\hat{\mathbf R}}$ and ${\hat{\mathbf P}}$, commutes
with ${\hat H}_{\rm I}(t)$. Therefore, the transformation generated by the
displacement operator does not change the internal state of the system; it only
acts on the center of mass variables.

\section{Separability of the spectrum}

The Hamiltonian (\ref{ham}) is time dependent because we have allowed for a
time dependent trap potential. Such a Hamiltonian does not possess true
stationary states. The most interesting case of a time-dependent harmonic
potential occurs when the time dependence is caused by a rotation of the trap.
Then, by going to the rotating frame we can eliminate the time dependence and
study the spectral properties of the resulting Hamiltonian. The price to be
paid for this simplification is the appearance of an extra term in the
Hamiltonian --- the scalar product $-{\bm\Omega}\!\cdot\!{\hat{\mathbf M}}$ of
the angular velocity ${\bm\Omega}$ and the total angular momentum of the system
${\hat{\mathbf M}}$,
\begin{eqnarray}\label{angmom}
 {\hat{\mathbf M}}
 = -i\int\!\!d^3r{\hat\psi}^\dagger({\mathbf r})({\mathbf r}\times{\mathbf\nabla})
 {\hat\psi}({\mathbf r}).
\end{eqnarray}
This term is responsible for the inertial forces: the Coriolis force and the
centrifugal force. The total Hamiltonian in the rotating frame is
\begin{eqnarray}
 {\hat H}_{\rm R} &=& {\hat H}_{\rm CM} + {\hat H}_{\rm I},\label{ham112}\\
 {\hat H}_{\rm CM} &=& \frac{{\hat{\mathbf P}}\!\cdot\!{\hat{\mathbf P}}}{2{\hat{N}}}
 + \frac{{\hat{\mathbf R}}\!\cdot\!{\hat{A}}\!\cdot\!{\hat{\mathbf R}}}{2{\hat{N}}}
 - {\bm\Omega}\!\cdot\!\frac{{\hat{\mathbf R}}\times{\hat{\mathbf P}}}{\hat{N}},
 \label{ham1cm}\\
 {\hat H}_{\rm I} &=& \frac{1}{2}\int\!\!d^3r{\hat\psi}^\dagger({\mathbf r})
 \left(-{\mathbf\nabla}_{\rm C}\!\cdot\!{\mathbf\nabla}_{\rm C}
 + {\mathbf r}_{\rm C}\!\cdot\!
 {\hat A}\!\cdot\!{\mathbf r}_{\rm C}\right){\hat\psi}({\mathbf r})\nonumber\\
 &+&\frac{1}{2}\!\int\!\!d^3r\!\!\int\!\!d^3r'{\hat\psi}^\dagger({\mathbf r})
 {\hat\psi}^\dagger({\mathbf r}') V({\mathbf r}-{\mathbf r}'){\hat\psi}({\mathbf r}')
 {\hat\psi}({\mathbf r})\nonumber\\
 &+& i{\bm\Omega}\!\cdot\!\int\!\!d^3r{\hat\psi}^\dagger({\mathbf r})
 ({\mathbf r}_{\rm C}\times{\mathbf\nabla}_{\rm C}){\hat\psi}({\mathbf
 r}),
\end{eqnarray}
where ${\hat{A}}$ is the time independent matrix whose eigenvalues $a_x$,
$a_y$, and $a_z$ are the squared frequencies of the trap potential. We use the
coordinate system with the axes directed along the principal directions of the
trap potential.

The transformation to the rotating frame preserves the separability of the
center of mass motion from the internal motion. Therefore, the spectrum of the
many-body Hamiltonian is a Cartesian product of the center of mass spectrum and
the spectrum of the internal motion. Each eigenvalue of the Hamiltonian ${\hat
H}_{\rm I}$ gives rise to the whole ladder made of the levels of the center of
mass Hamiltonian ${\hat H}_{\rm CM}$. These levels can be exactly calculated
regardless of the form of the mutual interaction. Since for a fixed number of
particles the commutation relations between ${\hat{\mathbf R}}$ and
${\hat{\mathbf P}}$ are canonical (up to a numerical factor of $n$), the center
of mass dynamics is that of a three-dimensional anisotropic harmonic oscillator
in a rotating frame. The characteristic frequencies of this oscillator can be
calculated from the following classical equations of motion determined by the
Hamiltonian (\ref{ham1cm}) (the equations of motion and the characteristic
frequencies given in Ref.~\cite{gpv} are incorrect)
\begin{eqnarray}\label{eqm}
 \frac{d^2{\mathbf R}(t)}{dt^2}
 = -{\hat{A}}\!\cdot\!{\mathbf R}(t)
 - {\bm\Omega}\times\left(2\frac{d{\mathbf R}(t)}{dt}
 + {\bm\Omega}\times{\mathbf R}(t)\right).
\end{eqnarray}
The characteristic equation for this system may be expressed in terms of six
invariant quantities built from $\hat{A}$ and ${\bm\Omega}$ and it has the form
\begin{widetext}
\begin{eqnarray}\label{charpol}
 &\omega^6 - \,\omega^4(2{\bm\Omega}^2+{\mathrm{Tr}}\{\hat{A}\})
 + \omega^2\left(({\bm\Omega}^2)^2
 +3{\bm\Omega}\!\cdot\!\hat{A}\!\cdot\!{\bm\Omega}
 - {\mathrm{Tr}}\{A\}\,{\bm\Omega}^2
 + {\mathrm{Tr}}\{A\}^2/2-{\mathrm{Tr}}\{\hat{A}^2\}/2\right)&\nonumber\\
 &- {\bm\Omega}^2\,{\bm\Omega}\!\cdot\!\hat{A}\!\cdot\!{\bm\Omega}
 + {\mathrm{Tr}}\{A\}\,{\bm\Omega}\!\cdot\!\hat{A}\!\cdot\!{\bm\Omega}
 -{{\bm\Omega}\!\cdot\!\hat{A}^2\!\cdot\!{\bm\Omega}
 - \mathrm{Det}}\{A\} = 0.&
\end{eqnarray}
\end{widetext}
One may solve this third order equation for $\omega^2$ but the formulas are
quite complicated. We give here the explicit solution only in the simple case,
when the angular rotation vector is directed along one of the principal axes,
say the $z$ axis, of the potential ellipsoid and the axes $x$ and $y$ are
directed along the two remaining principal directions. In this case the
frequency of oscillations along the $z$ direction is not modified by the
rotation, $\omega_z = \sqrt{a_z}$. The frequencies of the oscillations in the
perpendicular directions become equal to
\begin{eqnarray}
 \omega_{\pm} = \frac{\sqrt{2\Omega^2
 + a_+ \pm\sqrt{a_-^2 + 8\Omega^2a_+}}}{\sqrt{2}},
\end{eqnarray}
where $a_{\pm} = a_x \pm a_y$. The spectrum of ${\hat H}_{\rm CM}$ is discrete
when the frequencies $\omega_{\pm}$ are real. This is true (cf.~\cite{bmrs})
for slow rotations ($\Omega<\sqrt{a_x}$) or for fast rotations
($\Omega>\sqrt{a_y}$), where we have assumed for definitness that $a_x<a_y$. In
the intermediate region, the classical oscillations are unstable. This means
that for the intermediate values of the rotational frequency the spectrum of
${\hat H}_{\rm CM}$ is continuous and the condensate as a whole falls out of
the trap. The counterintuitive result that for fast rotations the oscillations
are stable is due to the action of the Coriolis force. The stabilizing effect
of the Coriolis force is well known in other physical situations. It plays the
crucial role in the Paul \cite{paul} trap and also in the dynamics of
nonspreading electronic wave packets in Rydberg states, called Trojan states
\cite{bke,bb2}.

The appearance of an unstable region of rotational velocities between the stable regions
of slow and fast rotations is not limited to the special case discussed above, but is a
general property of the anisotropic rotating oscillator. This can be seen from the
properties of the free term in the characteristic equation (\ref{charpol}). This term
(taken with the minus sign) is always equal to the product of the squares of the three
eigenfrequencies. In the absence of rotation this term is equal to $a_xa_ya_z$. In turn,
for fast rotations this term is again positive since it is dominated by
${\bm\Omega}^2\,{\bm\Omega}\!\cdot\!\hat{A}\!\cdot\!{\bm\Omega}$. The signature of an
unstable behavior is the change of sign that indicates that one of the characteristic
frequencies becomes imaginary. In the Appendix we show that for an anisotropic trap this
change of sign always takes place. Therefore there always exists an intermediate region
of rotational velocities for which the center of mass escapes from the rotating trap.

\section{Conclusions}

We have made a full use of the fact that for a many-body system of interacting atoms in a
general harmonic potential the dynamics of the center of mass decouples from the internal
dynamics. In the time-dependent description, this leads to the appearance of families of
states generated by the action of the displacement operator. The members of each family
are labeled by the solutions of the equations of motion for a classical particle in the
trap. Each family is built on an internal state of the system. The internal states are
not affected by the displacement operator.

For rotating traps --- that are often used in BEC experiments --- one may employ the
time-independent description by using the rotating frame. In that case one may study the
spectrum characterizing {\em stationary} states of the system. The decoupling of the
center of mass Hamiltonian leads to the splitting of the spectrum. The spectrum of the
full Hamiltonian is a Cartesian product of the center of mass spectrum and the spectrum
of the internal motion. The spectrum of the center of mass Hamiltonian may be discrete or
continuous depending on the value of the rotational velocity. For sufficiently slow
rotations the spectrum is discrete because the centrifugal force is not strong enough to
overcome the trap attraction. For sufficiently fast rotations the spectrum is again
discrete because the Coriolis force stabilizes the center of mass motion. In the two
stable regions the eigenvalue spectrum forms a triplet of ladders with equally spaced
rungs built on each eigenvalue of the internal Hamiltonian. The spacings are given by the
three frequencies, solutions of the characteristic equation (\ref{charpol}). For
intermediate values of the rotational velocity the spectrum is continuous that means that
the center of mass motion is unbounded --- the condensate escapes from the trap.

The spectrum of the center of mass Hamiltonian is not influenced by the
internal dynamics but the inverse is not true. The eigenvalue spectrum of the
internal Hamiltonian does depend on the properties of the trap. For example, in
an exactly soluble model with harmonic mutual interactions studied previously
\cite{bb1}, the trap potential modifies the frequency of the quadrupole
oscillations. In the present paper we have considered the case of arbitrary
two-body mutual interactions but our results are also valid in the general case
of $n$-body interactions provided the interaction potential is a function of
coordinate differences only.

The decoupling of the center of mass motion may have direct observational
consequences for small condensates, i.e. for condensates whose physical
dimensions are small as compared to the extension of the center of mass wave
function. Such condensates may be produced, for example, when interatomic
forces are attractive (as in lithium). Then, all many-particle correlation
functions are highly peaked at small distances between the particles. In this
case, the center of mass wave function contains significant (probabilistic)
information about the position of the {\em whole} condensate. In particular,
the vortex lines embedded in this wave function will have experimentally
testable consequences; the probability of finding the atoms of the condensate
close to the vortex line is small. Since the center of mass wave function obeys
the one-particle Schr\"odinger equation in a trap, the explicit solutions of
this equation exhibiting various vortex structures that were studied in detail
in Refs. \cite{bmrs} and \cite{bbs} will become relevant.

\begin{acknowledgements}

We would like to thank Mariusz Gajda for fruitful discussions.

\end{acknowledgements}

\appendix*
\section{}

In this Appendix we prove that in a rotating anisotropic trap there always
exists a range of rotational velocities ${\bm\Omega}$ where the center of mass
motion is unstable. In such a region one of the characteristic frequencies will
be imaginary. As mentioned in the main text, this proof is based on the
analysis of the characteristic equation (\ref{charpol}). The (minus) free term
in this equation is equal to the product of the three characteristic
frequencies squared. As seen from (\ref{charpol}), it is a biquadratic
polynomial in the absolute value of ${\bm\Omega}$
\begin{eqnarray}\label{charpol1}
 &&\omega_1^2\omega_2^2\omega_3^2
 = \Omega^4\,{\bm n}\!\cdot\!\hat{A}\!\cdot\!{\bm n}\nonumber\\
 &&- \Omega^2\left({\mathrm{Tr}}\{A\}\,{\bm n}\!\cdot\!\hat{A}\!\cdot\!{\bm n}
 -{\bm n}\!\cdot\!\hat{A}^2\!\cdot\!{\bm n}\right)
 + {\mathrm{Det}}\{A\},
\end{eqnarray}
where ${\bm n}$ is the unit vector in the direction of ${\bm\Omega}$. The signature of
the transition to an unstable region is the change of sign of this polynomial occurring
at its zeros. The polynomial (\ref{charpol1}) has zeros if the discriminant $\Delta$ is
positive
\begin{eqnarray}\label{delta}
 \Delta &=& \left({\mathrm{Tr}}\{A\}\,{\bm n}\!\cdot\!\hat{A}\!\cdot\!{\bm
 n}
 -{\bm n}\!\cdot\!\hat{A}^2\!\cdot\!{\bm n}\right)^2\nonumber\\
 &-&4{\mathrm{Det}}\{A\}{\bm n}\!\cdot\!\hat{A}\!\cdot\!{\bm n}.
\end{eqnarray}
Without loss of generality, we may assume that $a_x<a_y<a_z$. Then, by
rearranging the terms, we can easily obtain the form of $\Delta$ that exhibits
its positivity
\begin{eqnarray}\label{delta1}
 \Delta \!&=&\!\!\left(n_x^2a_x(a_z-a_y)+n_y^2a_y(a_z-a_x)+n_z^2a_z(a_x-a_y)\right)^2
 \nonumber\\
 &+& 4n_y^2n_z^2a_ya_z(a_z-a_x)(a_y-a_x)\geq 0.
\end{eqnarray}
The vanishing of $\Delta$ means that there is one double root for $\Omega^2$. In that
case there is no region of instability. This occurs only when the trap is not fully
anisotropic but has a symmetry axis and rotates around this direction.


\begin{thebibliography}{22}
\bibitem{gpv} J. J. Garc\'{\i}a-Ripoll, V. M. P\'erez-Garc\'{\i}a,
and V. Vekslerchik, Phys. Rev. E {\bf 64}, 056602 (2001).
\bibitem{pp} C. J. Pethick and L. P. Pitaevskii, Physical Review A {\bf 62},
033609 (2000).
\bibitem{glauber} R. Glauber, Phys. Rev. {\bf 131}, 2766 (1963).
\bibitem{scully} M. O. Scully and M. S. Zubairy, {\em Quantum Optics}, Cambridge
University Press, Cambridge, 1997.
\bibitem{bb} I. Bialynicki-Birula, Lett. in Math. Phys. {\bf 10}, 189 (1985).
\bibitem{bb1} Z. Bialynicka-Birula and I. Bialynicki-Birula,
Phys. Rev. A {\bf 33}, 1671 (1986).
\bibitem{paul} H. Paul, Rev. Mod. Phys. {\bf 62}, 531 (1990).
\bibitem{bke} I. Bialynicki-Birula, M. Kalinski,
and J. H. Eberly, Phys. Rev. Lett. {\bf 73}, 1777 (1994).
\bibitem{bb2} I. Bialynicki-Birula and Z. Bialynicka-Birula,
Phys. Rev. Lett. {\bf 77}, 4298 (1996).
\bibitem{bmrs} I. Bialynicki-Birula, T. M{\l}oduchowski, T. Rado\.zycki,
and C. \'Sliwa, Acta Phys. Polon. A {\bf 100} (Supplement), 29 (2001).
\bibitem{bbs} I. Bialynicki-Birula, Z. Bialynicka-Birula, and C. \'Sliwa, Phys.
Rev. A {\bf 61} 032110 (2000).
\end{thebibliography}
\end{document}